\documentclass[a4paper,11pt]{article}
\pdfoutput=1 % if your are submitting a pdflatex (i.e. if you have
\usepackage{jheppub} % for details on the use of the package, please
\usepackage{jheppub}
\usepackage[T1]{fontenc} % if needed
\usepackage[toc,page]{appendix}
\usepackage{dsfont}

\newcommand{\be}{\begin{equation}}
\newcommand{\ee}{\end{equation}}
\newcommand{\ba}{\begin{array}}
\newcommand{\ea}{\end{array}}
\newcommand{\bea}{\begin{eqnarray}}
\newcommand{\eea}{\end{eqnarray}}

\usepackage{ulem,fancyvrb}
\usepackage{xcolor}

\title{Two-component millicharged dark matter and  
the EDGES 21cm signal}

\author[a]{Qiaodan Li,}
\author[a]{Zuowei Liu} %,\note{Corresponding author.}}

\affiliation[a]{Department of Physics, Nanjing University, Nanjing 210093, China}

\abstract{

We propose a two-component dark matter explanation 
to the EDGES 21 cm anomalous signal. 
The heavier dark matter component is long-lived 
whose decay is primarily responsible for the 
relic abundance of the lighter dark matter which is 
millicharged. 
To evade the constraints from CMB, underground dark matter 
direct detection, and XQC experiments, 
the lifetime of the heavier dark matter 
has to be larger than $0.1\, \tau_U$, where 
$\tau_U$ is the age of the universe. 
Our model provides a viable realization of the 
millicharged dark matter model to explain the 
EDGES 21 cm, since the minimal model 
in which the relic density is generated via thermal 
freeze-out is ruled out by various constraints. 
}

\begin{document} 
\maketitle
\flushbottom

%\newpage
%~\newpage
%~\newpage

\section{Introduction}

The hyperfine transition of the cosmic hydrogen atom, 
which is known as the 21 cm signal, 
provides a powerful probe to the physics 
in the early universe; 
for reviews, see e.g., \cite{Furlanetto:2006jb} 
\cite{Morales:2009gs}
\cite{Pritchard:2011xb}.
Recently, the Experiment to Detect the Global Epoch of 
Reionization Signature (EDGES) has reported a 
new measurement on the sky-averaged radio spectrum 
centered at $\nu=78$ MHz, which corresponds to the 
21 cm absorption signal of the primordial hydrogen gas  
at redshift $z\simeq 17$ 
\cite{Bowman:2018yin}. 
The {differential} brightness temperature 
$T_{21}$ measured by the 
EDGES experiment at redshift $z\simeq17$ is 
\cite{Bowman:2018yin}
\be
T_{21} =-500^{+200}_{-500}~ {\rm mK}, 
\label{eq:-500mK}
\ee
which is about a factor of two larger than the value 
expected in the standard cosmology 
\cite{Bowman:2018yin, Cohen:2016jbh}. 
This hints that either the cosmic microwave background 
(CMB) could be hotter than expected 
\cite{Feng:2018rje, Fraser:2018acy, 
Moroi:2018vci, Kovetz:2018zes}, 
or the hydrogen gas could be colder than 
expected \cite{Barkana:2018qrx, Liu:2019knx, Creque-Sarbinowski:2019mcm, 
Munoz:2018pzp, Berlin:2018sjs, Barkana:2018lgd, Boddy:2018wzy, 
Jia:2018csj, 
Houston:2018vrf, 
Sikivie:2018tml, 
Fialkov:2018xre, Li:2018kzs, Kovetz:2018zan, Slatyer:2018aqg}. 

If the EDGES measurement is interpreted 
as caused by a colder gas, the question 
is how to cool the hydrogen gas? 
One simple solution is that if there exits 
some sort of particle interactions between 
dark matter (DM) and hydrogen atom, 
the primordial hydrogen 
gas can be cooled by DM, 
which is colder than gas. 
However, there exist very strong constraints 
from the CMB measurements as well as 
from other early universe measurements 
on DM interactions to standard model particles.  
Millicharged DM is one of the leading DM candidates  
to explain the EDGES anomaly because 
its interaction with baryons 
is proportional to $v^{-4}$, where $v$ 
is the relative velocity, and thus leads 
to a much smaller interaction cross 
section in the early universe 
than that needed at $z\simeq 17$ 
for the EDGES interpretation so that 
the early universe constraints 
can be significantly alleviated. 
Millicharged DM have been extensively investigated 
for the interpretation of the EDGES anomaly  
\cite{Barkana:2018qrx, 
Liu:2019knx, 
Creque-Sarbinowski:2019mcm, 
Munoz:2018pzp, 
Berlin:2018sjs, 
Barkana:2018lgd, Boddy:2018wzy, 
Fialkov:2018xre, Kovetz:2018zan, Slatyer:2018aqg}. 
The parameter space of the millicharged DM 
is constrained by various experiments, 
including accelerator experiments 
\cite{Prinz:1998ua}, 
the CMB anisotropy 
\cite{Dubovsky:2003yn} 
\cite{Dolgov:2013una} 
\cite{dePutter:2018xte}  
\cite{Kovetz:2018zan} 
\cite{Boddy:2018wzy}, 
the SN1987A  
\cite{Chang:2018rso}, 
and the Big Bang Nucleosynthesis
(BBN) 
\cite{Boehm:2013jpa}
\cite{Berlin:2018sjs}
\cite{Depta:2019lbe}; 
the allowed parameter space is that 
the millicharged DM mass is $0.1$ MeV $\lesssim 
m_\chi \lesssim 10$ MeV, the millicharge is 
$10^{-6} \lesssim Q/e \lesssim 10^{-4}$, and the 
mass fraction of the millicharged DM  is $0.0115\% 
\lesssim f \lesssim 0.4\%$. 
%%% 
However, as pointed out by 
Ref.\ \cite{Creque-Sarbinowski:2019mcm}, 
such a parameter space is ruled out by 
the $N_{\rm eff}$ limit with
the Planck 2018 data if the relic abundance 
of the millicharged DM is set by thermal freeze-out.
Recently, Ref.\ \cite{Liu:2019knx} proposed a new 
millicharged DM model in which the 
sub-component millicharged DM has a sizable  
interaction cross section 
with the other DM components 
so that the millicharged DM can be cooled  
by the other DM components; 
this reopens the parameter space 
that was previously excluded by 
various experimental constraints.

In this paper we propose a new DM model
that consists of two DM components: 
the lighter DM component is the millicharged DM, 
and the heavier DM component is unstable, 
which decays into the lighter DM component. 
In our model, the millicharged DM 
are primarily produced   
after the recombination so that 
the stringent constraints from 
CMB can be alleviated. 
We show that such a model can explain the 
21 cm anomaly observed by EDGES 
and satisfy various experimental constraints.  
%\zr{
The rest of the paper is organized as follows. 
We present our model 
in section \ref{sec:model}.
We compute the number density 
of the two DM components as a function of redshift 
in section \ref{sec:numberDensity}. 
The temperature change due to the heavier DM 
decays is derived 
in section \ref{sec:temperature:decay}.
We provide the time evolution equations 
for four different physics quantities 
in section \ref{sec:timeEvolutionEquations}.
The results of our numerical analysis is 
given 
in section \ref{sec:results}. 
We summarize our findings 
in section \ref{sec:conclusions}.
%}

%\newpage 

%%
%%

\section{The model}
\label{sec:model}

We extend the standard model (SM) by introducing a hidden sector 
that consists of three $U(1)$ gauge bosons, 
$X^i_\mu$ ($i=1, 2, 3$), and one Dirac 
fermion $\chi$ that is charged under both  
$X^2_\mu$ and $X^3_\mu$ gauge bosons. 
We use the Stueckelberg mechanism 
\cite{Kors:2005uz, 
Feldman:2006ce, 
Feldman:2006wb, 
Feldman:2007wj, 
Feldman:2009wv, 
Du:2019mlc} 
to provide mass to the three $U(1)$ gauge bosons; 
the new Lagrangian is given by 
\bea
\Delta {\cal L} =
 &-& \sum_{i=1,2,3} {1\over 4} X_{i \mu \nu} 
 X_{i}^{\mu \nu}
 +
 \bar \chi (i\gamma^\mu \partial_\mu - m_\chi) \chi 
 + g^\chi_2 X_{2}^{\mu} \bar \chi \gamma^\mu \chi
 + g^\chi_3 X_{3}^{\mu} \bar \chi \gamma^\mu \chi
  \nonumber \\
&-& {1 \over 2} 
( \partial_\mu \sigma_1 
+  m_{1}^{\prime} X_{1}^{\mu})^2 
- {1 \over 2} 
( \partial_\mu \sigma_2 
+  m_{1} X_{1}^{\mu}
+  m_{2}  X_{2}^{\mu} )^2 
  \nonumber \\
&-& {1 \over 2} ( 
\partial_\mu \sigma_3 
+ m_{3} X_{3}^{\mu} 
+ m_{4} B^{\mu} )^2, 
\label{eq:eL}
\eea
where $\sigma_1$, $\sigma_2$, and  $\sigma_3$ 
are the axion fields 
in the Stueckelberg mechanism, 
$B_\mu$ is the SM hypercharge boson, 
$m_\chi$ is the dark fermion mass, 
$g^\chi_2$ and $g^\chi_3$ are the gauge couplings,  
and $m_1^{\prime}$, $m_1$, $m_2$, 
$m_3$, and $m_4$ 
are the Stueckelberg mass terms.

After the spontaneous symmetry breaking in the SM, 
the mass matrix of  
the neutral gauge bosons
in the basis $(X_1,X_2, X_3, B, A^3)$, 
where $A^3$ is the third component of 
the $SU(2)_L$ gauge bosons in the SM, 
is given by
\be
M^2 = 
\begin{pmatrix} 
m_1^{\prime 2} + m_1^2  & m_1 m_2 & 0 & 0 &0 \cr
m_1 m_2 &  m_2^2 & 0 & 0 &0 \cr 
0 & 0 & m_3^2 & m_3 m_4 & 0 \cr
0 & 0 & m_3 m_4 & m_4^2 
+ {g_Y^2 v^2/4} & - {g_Y g_2 v^2/4}  \cr
0 & 0 &  0 & - {g_Y g_2 v^2/4} & {g_2^2 v^2/4}
\end{pmatrix}, 
\label{eq:massMatrix}
\ee
where $v$ is the vacuum expectation value 
of the SM Higgs, 
and $g_2$ and $g_Y$ are the $SU(2)_L$ 
and $U(1)_Y$ gauge couplings in the SM respectively. 
The mass matrix has a vanishing determinant 
such that there exists a massless mode to be 
identified as the SM photon.  
Because the mass matrix is block-diagonal, 
one can diagonalize the first two gauge bosons 
and the last three gauge bosons separately.

The mass matrix of the first two gauge bosons 
(the upper-left two by two block matrix in 
Eq.\ (\ref{eq:massMatrix}))
can be diagonalized 
via a rotation matrix ${\cal R}$ which is parameterized by 
a single angle $\theta$ 
\be
{\cal R} = 
\begin{pmatrix} 
\cos\theta & {{-}\sin\theta}  \cr
{\sin\theta} & \cos\theta  
\end{pmatrix}. 
\ee
The mass eigenstates, 
$Z_1$ and $Z_2$, 
are related to the gauge states via 
$Z_i={\cal R}_{ji} X_{j}$. 
{The rotation matrix ${\cal R}$ 
leads to an interaction between 
$\chi$ and $Z_1$ such that 
${\cal L}_{\rm int}  = \sin\theta g^\chi_2 
Z_{1}^{\mu} \bar \chi \gamma^\mu \chi 
\equiv v_{Z_1}^\chi 
Z_{1}^{\mu} \bar \chi \gamma^\mu \chi $.
We are interested 
in the parameter space where 
$\theta \ll 1$ 
so that $Z_1 \sim X_1$ and 
$Z_2 \sim X_2$. 
In our analysis, 
we take  
$m_1^\prime \sim 2 m_\chi$, 
$m_2 < m_\chi$, 
and $m_1 \ll m_2$ 
so that the two mass eigenstates 
have masses  
$m_{Z_1} \simeq m_1^\prime $ 
and 
$m_{Z_2} \simeq m_2$, 
and the mixing angle $\theta$ is 
given by 
$\theta \simeq m_1 m_2 /(m_1^{\prime 2}-m_2^2)$.}

The mass matrix of the last three gauge bosons 
(the bottom-right three by three block matrix 
in Eq.\ (\ref{eq:massMatrix})) 
can be diagonalized 
by an orthogonal matrix ${\cal O}$ such 
that $E_i={\cal O}_{ji} G_{j}$, where 
$G_j=(X_3, B, A^3)$ are the gauge states, 
and $E_i = (Z', Z, \gamma)$ are the mass eigenstates. 
Here 
$\gamma$ is the photon, 
$Z$ is the neutral gauge boson in the weak interaction, 
and $Z'$ is the  
new massive vector boson. 
Thus we have ${\cal O}^T M_{3\times 3}^2 {\cal O} = 
{\rm diag} (m^2_{Z'}, m^2_{Z}, 0)$, 
where 
$M_{3\times 3}^2$ is the bottom-right  
three by three block matrix in Eq.\ (\ref{eq:massMatrix}). 
Such a matrix diagonalization also leads to 
interactions between 
matter fields (both hidden sector fermion $\chi$ 
and SM fermions $f$) and the three mass 
eigenstates ($\gamma$, $Z$, and $Z'$). 
The interaction Lagrangian can be 
parameterized as follows 
\be
\bar f \gamma_\mu (v^f_i - \gamma_5 a^f_i) f E^\mu_i 
+ v^\chi_i \bar \chi \gamma_\mu  \chi E^\mu_i, 
\label{eq:LI}
\ee
where the vector and axial-vector couplings are given by 
\bea
v^f_i &=& (g_2 {\cal O}_{3i} - g_Y {\cal O}_{2i})T^{3 }_f/2
+ g_Y{\cal O}_{2i}Q_f , \\
\label{eq:cp1}
a^f_i &=& (g_2 {\cal O}_{3i} - g_Y{\cal O}_{2i})T^{3 }_f/2, \\
\label{eq:cp2}
v^\chi_i &=& g_{\chi} {\cal O}_{1i}. 
\label{eq:cp3}
\eea
Here $Q_f$ is the electric charge of the SM fermion, 
and $T_f^3$ is the quantum number of the left-hand 
chiral component {under SU(2)$_L$}.

{
Thus, the hidden sector fermion $\chi$ has a vector current 
interaction with the SM photon,  
\be
v_3^\chi A_\mu \bar \chi \gamma^\mu \chi 
\equiv  
\epsilon e A_\mu \bar \chi \gamma^\mu \chi, 
\ee
where we have defined an electric charge 
$\epsilon$ for the $\chi$ particle.  
In our analysis, we adopt the following 
model parameters: 
$m_3=100$ TeV, and 
$m_4/m_3 \ll 1$. 
In this case, the electric charge $\epsilon$ is given by 
{$\epsilon \simeq - (m_4/m_3) \cos\theta_W (g_3^\chi /e)$, 
where 
$\theta_W$ is the weak rotation angle in the SM.}
Since $m_4/m_3 \ll 1$ in our analysis, 
we have $\epsilon \ll 1$, 
which is often  
referred to as millicharge. 
$\chi$ is then the millicharged particle.}

\section{Two DM components}% 
\label{sec:numberDensity}

There are two DM particles in the hidden sector, 
the $Z_1$ boson and the hidden Dirac fermion 
$\chi$. 
In the very early universe,  
$Z_1$ is the dominant DM component, 
which is assumed to be nonthermally produced. 
The $Z_1$ DM component is long-lived and decays into $\bar \chi \chi$. 
The decay width of the $Z_1$ boson 
is given by 
\be
\Gamma (Z_1\rightarrow\bar\chi\chi) 
=\frac{m_{Z_1}}{12\pi}
\sqrt{1-4\frac{m_\chi^2}{m_{Z_1}^2}} 
\left( 1+2\frac{m_\chi^2}{m_{Z_1}^2} \right) 
(v_{Z_1}^\chi)^2
\simeq {\sqrt{m_{Z_1} \, \Delta m } \over 4 \sqrt{2} \pi}
\left( {g^\chi_2 \theta} \right)^2, 
\ee
where 
$v_{Z_1}^\chi = g^\chi_2 \sin \theta \simeq g^\chi_2 \theta$, 
$\Delta m \equiv m_{Z_1} - 2 m_\chi$, 
and we have assumed  
$\Delta m \ll m_{Z_1}$. 
In our analysis we have 
$\theta \ll 1$ and $\Delta m \ll m_{Z_1}$ 
so that $Z_1$ is long-lived with a lifetime  
\be
{\tau (Z_1) 
\simeq 
\sqrt{ {\rm MeV} \over m_{Z_1}}
\sqrt{ {\rm meV} \over \Delta m} 
\left( { 2.3 \times 10^{-16} \over g^\chi_2 \theta } \right)^2 \, 
 \tau_{17}, }
\ee
where $ \tau_{17} \sim 7 \times 10^{15}$ second   
is the time between the early universe and 
$z=17$.

The value of 
$\Delta m$ cannot be very large, 
otherwise the DM $\chi$ is 
significantly heated by the decay 
process $Z_1 \to \bar \chi \chi$ 
so that $\chi$ is unable   
to cool the baryons.  
The velocity of the $\chi$ particle 
is $v_\chi \simeq \sqrt{\Delta m/m_\chi}$ 
in the rest frame of $Z_1$, 
under the assumption of $\Delta m \ll m_\chi$. 
For the case where {$m_\chi \sim 100$ MeV  
and $\Delta m \sim {\cal O}$(meV), one has 
$v_\chi \sim 3.2 \times 10^{-6}$.}
Thus, in our analysis, we assume a sufficiently small 
mass difference,  
$\Delta m \sim {\cal O}$(meV), 
such that the $Z_1$ decay does not 
heat the $\chi$ DM significantly.

The $\chi$ DM component is mainly produced via the decay 
process $Z_1 \to \bar \chi \chi$ in the universe. 
We assume that the initial number density of $\chi$ is negligible. 
{The relic density of $\chi$ can also be produced via 
thermal freeze-out. 
There are two processes that contribute to the 
$\chi$ DM annihilation cross section: 
$\bar \chi \chi \to \gamma \to \bar f f$ 
and $\bar \chi \chi \to Z_2 Z_2$. 
In our analysis, 
$\sigma(\bar \chi \chi \to Z_2 Z_2) \gg 
\sigma(\bar \chi \chi \to \bar f f)$. 
The $\bar \chi \chi$ annihilation cross section into 
on-shell $Z_2$ bosons is given by \cite{Cline:2014dwa}
\be
\langle \sigma v \rangle (\bar \chi \chi \to Z_2 Z_2) 
\simeq { (g_2^\chi)^4 \over 16  \pi m_\chi^2}
{(1-r^2)^{3/2} \over (1-r^2/2)^{2}}, 
\ee
where $r=m_{Z_2}/m_\chi$. 
For the case where $g_2^\chi = 1$, 
$r=1/2$, 
and $m_\chi = 5$ MeV, 
one has 
$\langle \sigma v \rangle (\bar \chi \chi \to Z_2 Z_2)
\simeq 0.3$ barn, leading to a mass fraction  
as $f_\chi \sim 10^{-11}$. 
Thus, the contribution to the relic density of $\chi$ 
from thermal freeze-out is negligible, 
and the constraints imposed on thermal freeze-out 
millicharged DM (e.g.\ Ref.\ \cite{Creque-Sarbinowski:2019mcm}) 
are not 
directly applicable to our model. 
}

The total number of $Z_1$ 
particles in a comoving volume at time $t$ is given by 
\be
N_{Z_1}(t) = N_{Z_1} (0) e^{ - t / \tau } , 
\ee
where 
$\tau$ is the lifetime of the $Z_1$ particle, 
and $N_{Z_1} (0)$ is the total number of $Z_1$  
particles at time $t=0$. 
In our analysis, we set 
$t=0$ at redshift $z_0=10^6$. 
The number of $\chi$ particles at time $t$ 
in a comoving volume is 
$
N_\chi (t) = 2N_{Z_1}(0) - 2N_{Z_1}(t). 
$
Thus, the number density of $\chi$ is related 
to the number density of $Z_1$ via 
\be
n_\chi   = 2( e^{t/\tau}-1)   n_{Z_1} .
\ee
where $n_{Z_1}\ (n_\chi)$ is the number density of 
the ${Z_1}\ (\chi)$ particle. 
In our analysis $m_{Z_1} \simeq 2 m_\chi$, 
so the mass fraction of the millicharged DM 
$\chi$ at redshift $z$ in the total DM 
 is given by 
\be
f_\chi (z) \simeq  (1-e^{-t(z)/\tau}), 
\ee
where $t(z)$ is the time between early universe (which we take 
to be $z_0=10^6$) and redshift $z$.\footnote{The formulas to 
compute $t(z)$ are given in Appendix \ref{section:time}.} 
Fig.\ (\ref{fig:fraction}) shows the mass 
fraction of $\chi$ as a function of the redshift 
$z$ for different lifetimes $\tau$. 
CMB observations provide strong constraints 
on millicharged DM; only 0.4\% DM can be millicharged 
unless the millicharge is negligible 
\cite{dePutter:2018xte}  
\cite{Kovetz:2018zan} 
\cite{Boddy:2018wzy}. 
This leads to an lower bound on the $\tau$, 
which is $\sim 3.6 \times 10^{15}$ s.

\begin{figure}[htbp]
\begin{centering}
\includegraphics[width=0.5\textwidth]{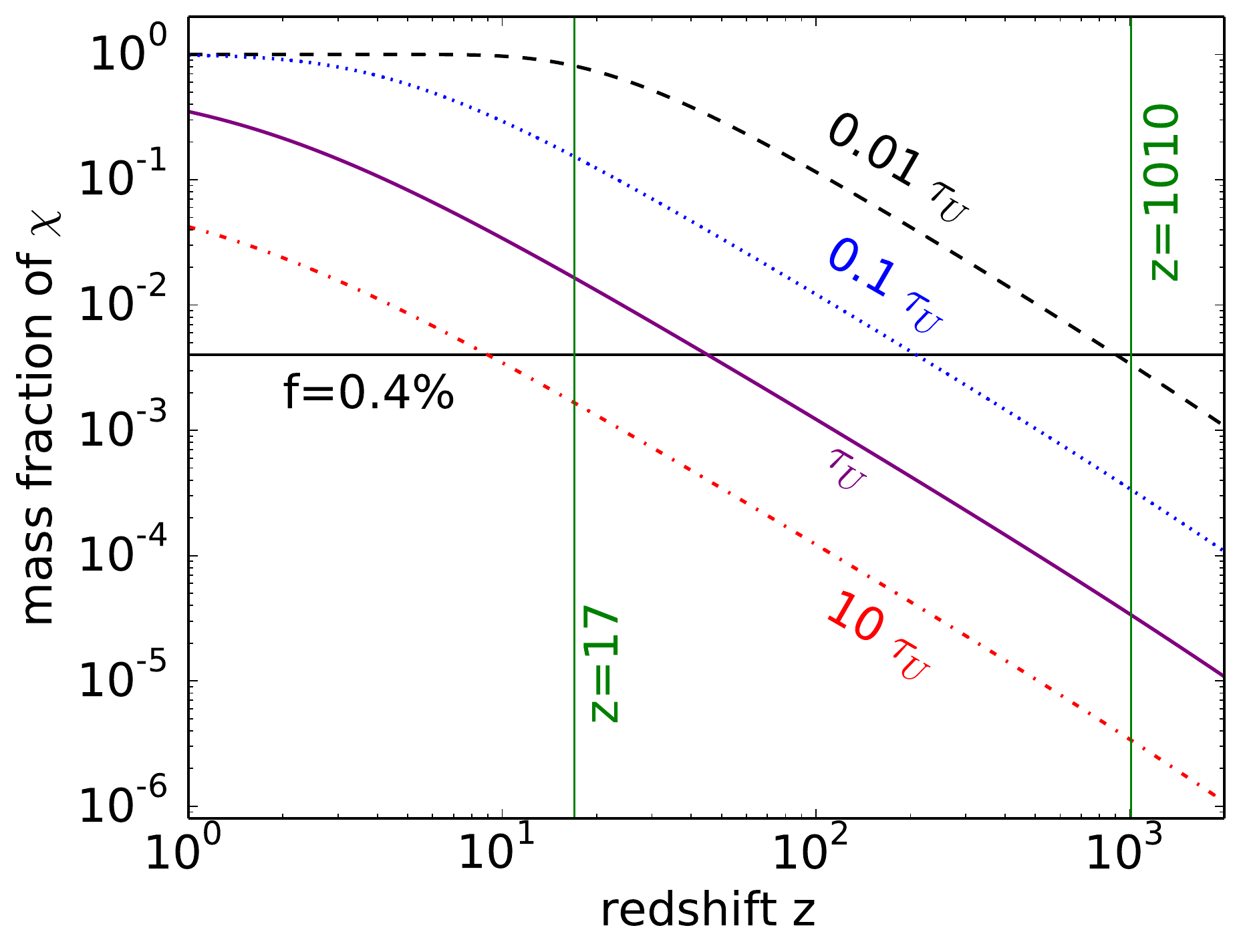}
\caption{Mass fraction of $\chi$ in the total 
DM as a function of the redshift $z$. 
$\tau_U \simeq 4.3 \times 10^{17}$ s is the age of the 
universe.}
\label{fig:fraction}
\end{centering}
\end{figure}

Because 
$\Delta m \ll m_\chi$, the total DM density $\rho_{Z_1} + \rho_\chi$ 
at redshift $z$ is given by  
$
\rho_{Z_1} + \rho_\chi =  \rho_{\rm DM, 0} \, (1+z)^3. 
$
Thus the number density of $\chi$ particles at redshift $z$ is 
given by 
\be
n_\chi (z) =  {\rho_{\rm DM,0} \over m_\chi} \,
  (1+z)^3 \, f_\chi(z) \simeq 
  {\rho_{\rm DM,0} \over m_\chi} \,
  (1+z)^3 \, (1-e^{-t(z)/\tau}), 
  \label{eq:nx}
\ee
where $\rho_{\rm DM, 0} = \Omega_{\rm DM} \, \rho_{\rm cr,0}$ 
is the current DM density where 
$\rho_{\rm cr,0}=1.054 h^2\times 10^4$ 
eV cm$^{-3}$ is the critical density \cite{Kolb:1990vq}. 
In our analysis, 
we use $\Omega_{\rm DM} h^2 = 0.1186$ \cite{Ade:2015xua}.

\section{DM temperature increase generated by decays}
\label{sec:temperature:decay}

The DM $\chi$ is heated by  
the $Z_1 \to \chi \bar \chi$ decay process 
because of the difference between 
the $Z_1$ mass and twice of the $\chi$ mass. 
To compute this effect, consider the 
kinetic energy $\Delta q$ that 
goes into the $\chi \bar \chi$ final state 
for the decay process $Z_1 \to \chi\chi$  
\be
\Delta{q}=\Delta{m}+\frac{3}{2}k_BT_{Z_1}, 
\ee
where $3k_BT_{Z_1}/2$ is the averaged kinetic 
energy of $Z_1$ with $T_{Z_1}$ being the temperature 
of the $Z_1$ particle 
and $k_B$ being the Boltzmann constant. 
Here we have assumed that $Z_1$ is non-relativistic 
and $\Delta m$ is sufficiently small such that $\chi$ 
is also non-relativistic. 
The change of the particle numbers 
in the comoving volume per unit time due to decay 
are given by 
$
\dot N_{Z_1} = - \Gamma_{Z_1} N_{Z_1} 
$ and 
$
\dot N_\chi = -2 \dot N_{Z_1}, 
$
where the dot denotes the 
derivative with respect to time, 
$N_\chi$ ($N_{Z_1}$) is the particle number of 
$\chi$ ($Z_1$), and 
$\Gamma_{Z_1}$ is the decay width 
for the process $Z_1\rightarrow\chi\chi$. 
The total kinetic energy transfer to the $\chi$ particles 
per unit time 
from $Z_1$ decays is given by 
$
\Delta q |\dot N_{Z_1}|, 
$
which is equal to the change of the kinetic energy of  
the $\chi$ particles per unit time 
%$
%$
\be
\Delta q |\dot N_{Z_1}| 
=\frac{d}{dt}
\left( \frac{3}{2} k_B T_\chi N_\chi \right)
=\frac{3}{2} k_B (T_\chi  \dot N_\chi 
+N_\chi  \dot T_\chi). 
\ee
Thus, the $\chi$ temperature change per unit time 
due to $Z_1$ decays is given by 
\be
\dot T_\chi
= \frac{n_{Z_1}}{n_\chi} \left[ 
\frac{2}{3k_B} \Delta m
+T_{Z_1}  -2T_\chi \right]
\Gamma_{Z_1}. 
\label{eq:decayQterm}
\ee
%}

\section{Time evolution equations} 
\label{sec:timeEvolutionEquations}

To compute the baryon temperature at redshift $z=17$, 
we numerically solve the temperature evolutions 
for various quantities.

The time evolution equation of the baryon temperature $T_b$ 
is given by (see e.g.\ \cite{Munoz:2015bca, Munoz:2018pzp}) 
\be
\frac{dT_b}{dt} = -2HT_b+\frac{2}{3}\frac{dQ_b}{dt}
+\Gamma_C(T_\gamma-T_b),
\label{eq:TEE-bt}
\ee
where 
$H$ is the Hubble parameter,\footnote{See 
Appendix \ref{section:time} for the calculation of $H$.}  
$T_\gamma$ is the CMB temperature, 
$Q_b$ is the 
energy transfer term due to DM-baryon 
scatterings, and  
$\Gamma_C$ is the Compton scattering rate 
which describes the effects due to 
CMB-baryon interactions.  
The 
Compton scattering rate is given by \cite{Munoz:2018pzp} 
\be
\Gamma_C=\frac{8\sigma_T  x_e} 
{3(1+f_{\rm He})m_e} U
\ee
where $\sigma_T$ is 
the Thomson cross section, 
$f_{\rm He}$ is the Helium fraction, 
and $U$ is the energy density. 
In our analysis, we use 
$\sigma_T=6.65 \times 10^{-25}$ cm$^2$ \cite{Kolb:1990vq}, 
$f_{\rm He} = 0.08$ \cite{Munoz:2018pzp}, 
and $U=(\pi^2/15) \, T_\gamma^4 
= 0.26(1+z)^4\ \rm{eV/cm^3}$ 
 \cite{Seager:1999km} \cite{Peebles:1994xt}.

The time evolution equation of the temperature $T_\chi$ 
of the lighter DM component $\chi$  
is given by 
\be
\frac{dT_\chi}{dt} = -2HT_\chi
+\frac{2}{3}\frac{dQ_\chi}{dt}
+ 
{\frac{n_{Z_1}}{n_\chi} 
\left[ \frac{2}{3k_B} \Delta m
+T_{Z_1}-2 T_\chi \right] \Gamma_{Z_1}}, 
\label{eq:TEE-dmt}
\ee 
where 
$Q_\chi$ is the 
energy transfer term due the interaction 
between DM $\chi$ and baryons. 
The first two terms on the right-hand 
side of Eq.\ \eqref{eq:TEE-dmt} 
represent the effects due to 
universe expansion and the 
DM-baryon scattering respectively, 
which are similar to the first two terms 
in Eq.\ \eqref{eq:TEE-bt}. 
The third term on the right-hand side 
of Eq.\ \eqref{eq:TEE-dmt} is 
new and is 
due to the decay of the $Z_1$ particle 
in our model, as discussed in 
section \ref{sec:temperature:decay}.

In addition, we also solve 
the time evolution equation of  
the relative bulk velocity between 
baryon and DM, 
$V_{\chi b} = | \vec V_{\chi b} |$ where 
$\vec V_{\chi b} \equiv \vec V_{\chi} - \vec V_{b}$
\cite{Munoz:2015bca}, 
\be
\frac{dV_{\chi b}}{dt} = -HV_{\chi b}-D(V_{\chi b}), 
\label{eq:TEE-bv}
\ee 
and the time evolution equation of the 
ionization fraction $x_e \equiv n_e/n_H$ \cite{AliHaimoud:2010dx}
\be
\frac{dx_e}{dt} = -C \left[ n_H\alpha_B {x_e}^2
-4(1-x_e)\beta_B 
\exp\left( {3E_0 \over 4T_\gamma} \right)  
\right]. 
\label{eq:TEE-if}
\ee
We follow Ref.\ \cite{AliHaimoud:2010dx} 
to obtain the various coefficients in  
Eq.\ \eqref{eq:TEE-if}.
The formulas of $Q_b$, 
$Q_\chi$, and $D(V_{\chi b})$ 
for millicharged DM used in our analysis 
are given in 
Appendix \ref{section:mCP}.

\section{Results}
\label{sec:results}

\begin{figure}[htbp]
\begin{centering}
\includegraphics[width=0.8\textwidth]{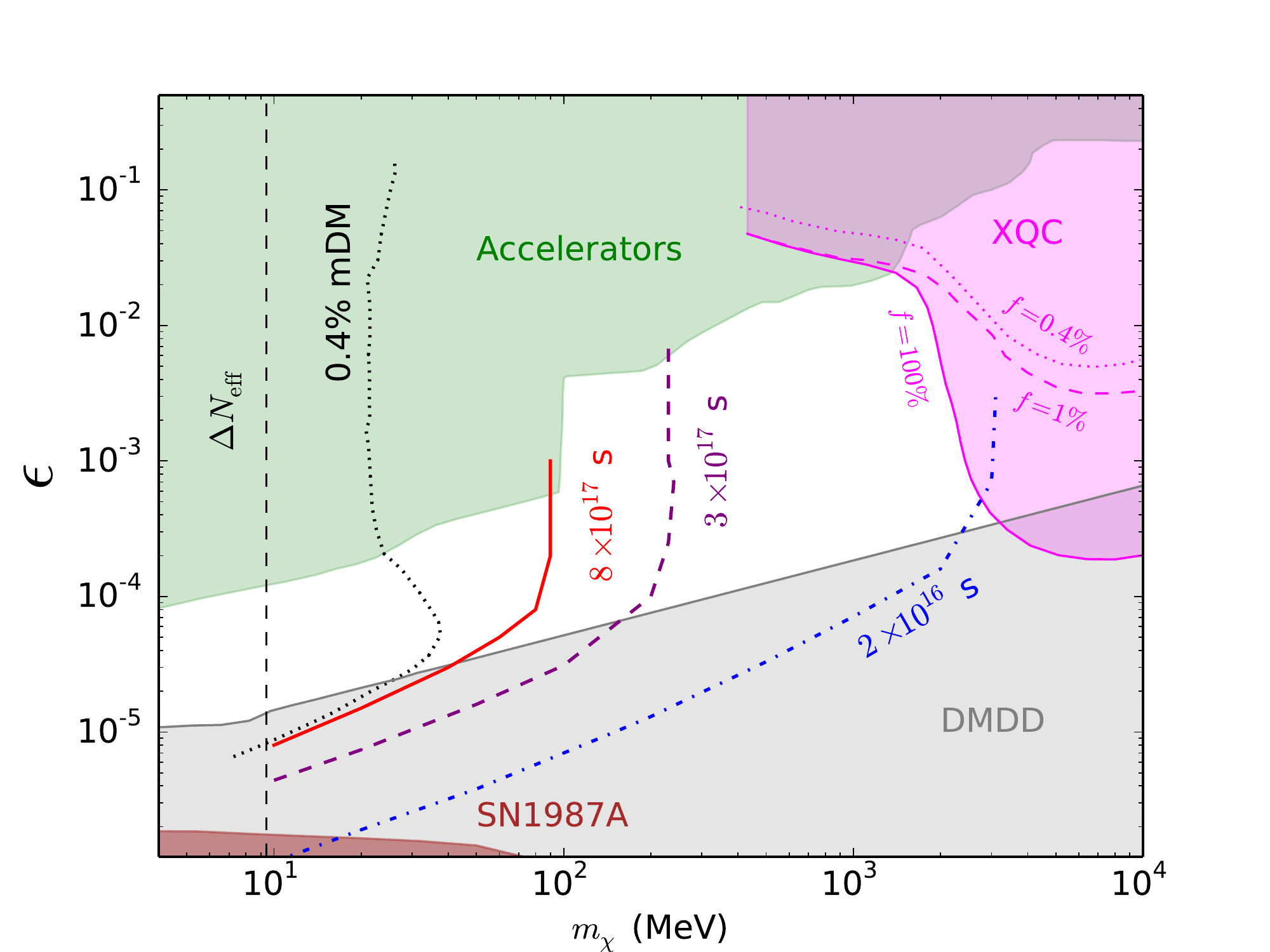}
\caption{Parameter space spanned by the millicharge $\epsilon$ 
and the DM mass $m_\chi$. 
Model points in which $T_b = 5$ K at $z=17$ 
correspond to three different lifetimes: 
$\tau=2\times 10^{16}$ s (blue dot-dashed), 
$\tau=3 \times 10^{17}$ s (purple dashed), 
and $\tau=8\times 10^{17}$ s (red solid), 
with the mass fraction of the 
millicharged DM component being 
$0.06\%$, $0.004\%$, and $0.001\%$ 
at $z=1100$,  
and 100\%, 77\%, and 42\% today respectively. 
%% Accelerators 
The green shaded region is excluded  
by various accelerator experiments, including 
SLAC electron beam dump \cite{Prinz:1998ua}, 
CMS \cite{CMS:2012xi}, 
MiniBooNE and LSND  
\cite{Magill:2018tbb}, 
ArgoNeuT \cite{ArgoNeuT:2019ckq}, 
milliQan demonstrator \cite{Ball:2020dnx},
and others  
\cite{Davidson:2000hf} \cite{Badertscher:2006fm}. 
%% DMDD 
The gray shaded region indicates the 
parameter region excluded by 
the dark matter direct detection 
(DMDD) experiments; 
above the DMDD region, 
millicharged DM is absorbed by 
the rocks on top of underground labs 
\cite{Emken:2019tni} 
\cite{Liu:2019knx}.
%% XQC 
The magenta region is ruled out  
by the rocket experiment XQC for 
mass fractions: 
$f=100\%$ (solid) \cite{Mahdawi:2018euy}, 
$f=1\%$ (solid) \cite{Mahdawi:2018euy},
and $f=0.4\%$ (solid) \cite{Emken:2019tni}.
The brown shaded region is  
excluded by the SN1987A data \cite{Fabbrichesi:2020wbt}. 
The black dashed vertical line indicates the upper bound 
on DM mass 
due to $\Delta N_{\rm eff}$ from CMB 
\cite{Boehm:2013jpa}
\cite{Berlin:2018sjs}. 
The black dotted line indicates the parameter 
space of the minimal millicharged DM model 
with a mass fraction of 0.4\% 
to explain EDGES data \cite{Liu:2019knx}. 
}
\label{fig:results}
\end{centering}
\end{figure}

We solve simultaneously the four time evolution equations  
for $T_b$, $T_\chi$, $V_{\chi b}$ and $x_e$ 
from redshift $z=1010$ to $z=10$. 
The baryon temperature $T_b$ at $z=1010$ is assumed 
to be equal to the CMB temperature 
$T_\gamma = T_0 \, (1+z)$ where $T_0 = 2.7$ K, 
since these two components are tightly coupled 
in the early universe. 
The temperatures for both DM components 
are assumed to be negligible in the early universe, 
so we set $T_{Z_1}= 0$ and $T_\chi = 0$ 
at $z=1010$. 
This is due to the fact that in our model 
$Z_1$ does not interact with the 
SM particles and $\Delta m \ll m_\chi$. 
We also set $x_e=0.05$ \cite{McGaugh:2018dfk} 
and $V_{\chi b}= 29$ km/s \cite{Munoz:2018pzp} 
at redshift $z=1010$.

\begin{table}[htbp]
\begin{center}
\setlength{\tabcolsep}{1mm}{
\begin{tabular}{|c|c|c|c|c|c|c|c||c|c|c|c|} 
\hline 
Model & $m'_1$ & $m_1$ & $m_2$ 
& $m_3$ & $m_4$ & $m_\chi$ 
& $\Delta m$ 
& $\theta$ & $\epsilon$  & $\tau_{Z_1} (s)$
\cr \hline
A & 160 & {$10^{-13}$} & {1.55} & $10^8$ & $2.76 \times 10^3$ & $\sim$80 
& $10^{-9}$  &  {$6.1 \times 10^{-18}$} & $8\times 10^{-5}$ & {$8 \times 10^{17}$} \cr \hline
$B$ & 400 & {$10^{-8}$} & {$1.25 \times 10^{-4}$} & $10^8$ & $3.45 \times 10^3$ & $\sim$200 
& $10^{-9}$  & {$7.8 \times 10^{-18}$} & $1 \times 10^{-4}$ & {$3 \times 10^{17}$}  \cr \hline
\end{tabular}}
\caption{Benchmark model points. 
All the masses are in unit of MeV.  
We take $g_2^\chi = g_3^\chi = 1$ in our analysis.}
\label{tab:benchmark}
\end{center}
\end{table}

We scan the parameter space spanned by 
${\epsilon}$ and $m_\chi$ for three 
different decay lifetimes of $Z_1$: 
$\tau=2 \times 10^{16}$ s, 
$3 \times 10^{17}$ s, and 
$8 \times 10^{17}$ s. 
In our analysis, we fix $\Delta m = 1$ meV.
In order to explain the EDGES data, the baryon temperature $T_b$ 
has to at least $5.1$ K at $z=17$ \cite{Liu:2019knx}. 
Fig.\ (\ref{fig:results}) shows 
the parameter space where the baryon temperature 
can be cooled to be 
$T_b=5.1$ K at $z=17$ for three different lifetimes of $Z_1$: 
$2 \times 10^{16}$ s (blue dot-dashed), 
$3 \times 10^{17}$ s (purple dashed), and 
$8 \times 10^{17}$ s (red solid). 
The mass fractions of the $\chi$ DM 
at $z=1100$ are about 
$0.06\%$, $0.004\%$, and $0.001\%$ 
for the lifetimes 
$\tau=2\times 10^{16}$ s, 
$\tau=3 \times 10^{17}$ s, 
and $\tau=8\times 10^{17}$ s respectively, 
which are smaller than $0.4$\% 
required by the CMB data 
\cite{dePutter:2018xte}, 
\cite{Kovetz:2018zan}, 
\cite{Boddy:2018wzy}. 
We find that the viable 
parameter in our model is much larger 
than the minimal millicharged DM 
model,\footnote{In the minimal millicharged DM 
model, the millicharge interaction is responsible for 
both DM thermal freeze-out and cooling 
of the cosmic hydrogen atoms.} 
which is indicated by 
the black dotted line 
with 0.4\% millicharged DM \cite{Liu:2019knx}. 
Various experimental constraints 
are considered in Fig.\ (\ref{fig:results}).
These include the underground 
dark matter direct detection 
(DMDD) experiments
\cite{Emken:2019tni}
\cite{Liu:2019knx}, 
the XQC experiment 
\cite{Mahdawi:2018euy}
\cite{Emken:2019tni}, 
SN1987A \cite{Fabbrichesi:2020wbt}, 
and the 
accelerator experiments: 
SLAC electron beam dump \cite{Prinz:1998ua}, 
CMS \cite{CMS:2012xi}, 
MiniBooNE \cite{Magill:2018tbb}, 
LSND \cite{Magill:2018tbb}, 
ArgoNeuT \cite{ArgoNeuT:2019ckq}, 
and milliQan demonstrator \cite{Ball:2020dnx}. 
Two benchmark model points that 
can explain the 21 cm 
anomaly while satisfying various constraints are 
presented in 
Table (\ref{tab:benchmark}). 
The mass fractions of the $\chi$ DM 
at today are  
100\%, 77\%, and 42\%
for the lifetimes 
$\tau=2\times 10^{16}$ s, 
$\tau=3 \times 10^{17}$ s, 
and $\tau=8\times 10^{17}$ s respectively. 
We find that the $Z_1$ lifetime 
$2 \times 10^{16}$ are nearly excluded by 
both the underground DMDD  and 
the XQC constraints. 
In order to evade the XQC constraints, 
the $Z_1$ lifetime   
has to be 
$\geq 2 \times 10^{16}$ s. 

\section{Conclusions}
\label{sec:conclusions}

We construct a new millicharged DM model to 
explain the recent 21 cm anomaly.  
In our model, the millicharged DM $\chi$ is 
a subcomponent in the early universe 
and is mainly produced via decays of 
the other DM component $Z_1$. 
The DM annihilation cross section 
$\bar \chi \chi \to Z_2 Z_2$ is so 
strong that the relic abundance due to 
thermal freeze-out is negligible. % 
We compute the heating term due to 
the decay process $Z_1 \to \bar \chi \chi$ 
and include it in our numerical calculations 
of the time evolution equation of the DM temperature. 
We find that the model can explain the 
EDGES 21 cm anomaly while satisfying  
various experimental constraints, 
including those from colliders, 
XQC, underground DMDD, and 
CMB.

\section{Acknowledgement}

We thank Ran Ding 
and Mingxuan Du  
for helpful discussions. 
The work is supported in part  
by the National Natural Science 
Foundation of China under Grant No.\ 
11775109.

\appendix

\section{Time}
\label{section:time}

The time $t(z)$ at redshift $z$ is given by 
\be
t(z)=\int_z^{z_0} {dz' \over H(z')(1+z')}, 
\ee
where $H$ is the Hubble parameter. 
Here we use ${z_0}=10^6$. 
We compute the Hubble parameter at redshift $z$ via 
\be
H(z)=H_0 \sqrt{\Omega_R(1+z)^4+\Omega_m(1+z)^3+\Omega_\Lambda}, 
\ee
where 
$H_0 \equiv 100h\ {\rm km\ s^{-1}
\ Mpc^{-1}}$ is the present Hubble parameter, 
$\Omega_R$, 
$\Omega_m$, 
and 
$\Omega_\Lambda$
are the density of radiation, 
matter, and dark energy respectively. 
In our analysis, we adopt the following values:  
$\Omega_R= 2.47\times10^{-5}/h^2$ \cite{Dodelson:2003ft}, 
$\Omega_m=0.308$, 
$\Omega_\Lambda=0.692$, 
and $h=0.678$ \cite{Ade:2015xua}.

\section{Millicharged DM formulas}
\label{section:mCP}

We provide the formulas 
of $\sigma_{0,t}$, $Q_b$, 
$Q_\chi$, and $D(V_{\chi b})$ 
for millicharged DM used in our analysis.

The scattering cross section between 
millicharged DM and baryons can be 
parameterized as  
$\sigma_t=\sigma_{0,t}v^{-4}$
where $v$ is the relative velocity between 
DM and baryons, 
and 
$
\sigma_{0,t} = 2 \pi \alpha^2 \epsilon^2 \xi / \mu_{\chi t}^2
$
where $\alpha$ is the fine structure constant, 
$\epsilon$ is the millicharge, 
$\mu_{\chi t}$ is the reduced mass of DM $\chi$ 
and the target particle $t$, 
and $\xi$ is the Debye logarithm
\cite{McDermott:2010pa} 
\cite{Munoz:2018pzp}
$
\xi = \ln \left[ 
{ 9 T_b^3 / (4 \pi \epsilon^2 \alpha^3 x_e n_H)}
\right].  
$

The baryon heating term due to interactions with 
millicharged DM is given by 
\cite{Munoz:2015bca}
\cite{Munoz:2018pzp}
\be
\frac{dQ_b}{dt}= n_\chi {x_e} \sum_{t=e,p}
\frac{m_t m_\chi }{(m_\chi+m_t)^2}
\frac{\sigma_{0,t}}{u_{{\rm th},t}}
\left[\sqrt{\frac{2}{\pi}}\frac{e^{-r_t^2/2}}{u_{{\rm th},t}^2}(T_\chi-T_b)
+m_\chi\frac{F(r_t)}{r_t}\right]. 
\ee
where {$u_{\rm th}^2 \equiv T_b/m_b+T_\chi/m_\chi$}, 
$r \equiv V_{\chi b}/u_{\rm th}$, and 
$
F(r) = {\rm erf}\left({r/\sqrt{2}} \right)-\sqrt{{2}{\pi}} r e^{-r^2/2}.
$ 
Here we assume that electron and proton share a common 
temperature $T_b$ 
with the hydrogen atom. 
The DM heating term due to interactions with 
baryons is given by 
\be
\frac{dQ_\chi}{dt}=n_H x_e\sum_{t=e,p}
\frac{m_\chi m_t}{(m_\chi+m_t)^2}
\frac{\sigma_{0,t}}{u_{{\rm th,}t}}
\left[\sqrt{\frac{2}{\pi}}\frac{e^{-r_t^2/2}}{u_{{\rm th,}t}^2}(T_b-T_\chi)
+m_t\frac{F(r_t)}{r_t}\right],
\ee
where $n_e = n_p = n_H x_e$ is assumed.

The  $D$ term in Eq.\ \eqref{eq:TEE-bv} is given by  \cite{Munoz:2015bca} 
\be
D(V_{\chi b}) = { \rho_m \sigma_0 \over m_\chi + m_b }
{ F(r) \over V_{\chi b}^2 }, 
\ee
where we consider both electron and proton as the target baryons. 


\begin{thebibliography}{99}

%\cite{Furlanetto:2006jb}
\bibitem{Furlanetto:2006jb}
S.~Furlanetto, S.~P.~Oh and F.~Briggs,
``Cosmology at Low Frequencies: The 21 cm Transition and the High-Redshift Universe,''
Phys. Rept. \textbf{433}, 181-301 (2006)
%doi:10.1016/j.physrep.2006.08.002
[arXiv:astro-ph/0608032 [astro-ph]].
%819 citations counted in INSPIRE as of 09 Aug 2021



%\cite{Morales:2009gs}
\bibitem{Morales:2009gs}
M.~F.~Morales and J.~S.~B.~Wyithe,
``Reionization and Cosmology with 21 cm Fluctuations,''
Ann. Rev. Astron. Astrophys. \textbf{48}, 127-171 (2010)
%%doi:10.1146/annurev-astro-081309-130936
[arXiv:0910.3010 [astro-ph.CO]].
%255 citations counted in INSPIRE as of 09 Aug 2021


%\cite{Pritchard:2011xb}
\bibitem{Pritchard:2011xb}
J.~R.~Pritchard and A.~Loeb,
``21-cm cosmology,''
Rept. Prog. Phys. \textbf{75}, 086901 (2012)
%doi:10.1088/0034-4885/75/8/086901
[arXiv:1109.6012 [astro-ph.CO]].
%431 citations counted in INSPIRE as of 08 Aug 2021


%\cite{Bowman:2018yin}
\bibitem{Bowman:2018yin}
J.~D.~Bowman, A.~E.~E.~Rogers, R.~A.~Monsalve, T.~J.~Mozdzen and N.~Mahesh,
``An absorption profile centred at 78 megahertz in the sky-averaged spectrum,''
Nature \textbf{555}, no.7694, 67-70 (2018)
%doi:10.1038/nature25792
[arXiv:1810.05912 [astro-ph.CO]].
%344 citations counted in INSPIRE as of 28 Apr 2020    year = "2018"


%\cite{Cohen:2016jbh}
\bibitem{Cohen:2016jbh}
A.~Cohen, A.~Fialkov, R.~Barkana and M.~Lotem,
``Charting the Parameter Space of the Global 21-cm Signal,''
Mon. Not. Roy. Astron. Soc. \textbf{472}, no.2, 1915-1931 (2017)
%doi:10.1093/mnras/stx2065
[arXiv:1609.02312 [astro-ph.CO]].
%98 citations counted in INSPIRE as of 09 Aug 2021


%\cite{Feng:2018rje}
\bibitem{Feng:2018rje}
C.~Feng and G.~Holder,
``Enhanced global signal of neutral hydrogen due to excess radiation at cosmic dawn,''
Astrophys. J. Lett. \textbf{858}, no.2, L17 (2018)
%doi:10.3847/2041-8213/aac0fe
[arXiv:1802.07432 [astro-ph.CO]].
%136 citations counted in INSPIRE as of 09 Aug 2021


%\cite{Fraser:2018acy}
\bibitem{Fraser:2018acy}
S.~Fraser, A.~Hektor, G.~H\"utsi, K.~Kannike, C.~Marzo, L.~Marzola, C.~Spethmann, A.~Racioppi, M.~Raidal and V.~Vaskonen, \textit{et al.}
``The EDGES 21 cm Anomaly and Properties of Dark Matter,''
Phys. Lett. B \textbf{785}, 159-164 (2018)
%doi:10.1016/j.physletb.2018.08.035
[arXiv:1803.03245 [hep-ph]].
%115 citations counted in INSPIRE as of 09 Aug 2021

%\cite{Moroi:2018vci}
\bibitem{Moroi:2018vci}
T.~Moroi, K.~Nakayama and Y.~Tang,
``Axion-photon conversion and effects on 21 cm observation,''
Phys. Lett. B \textbf{783}, 301-305 (2018)
%doi:10.1016/j.physletb.2018.07.002
[arXiv:1804.10378 [hep-ph]].
%50 citations counted in INSPIRE as of 27 Oct 2021

%\cite{Kovetz:2018zes}
\bibitem{Kovetz:2018zes}
E.~D.~Kovetz, I.~Cholis and D.~E.~Kaplan,
``Bounds on ultralight hidden-photon dark matter from observation of the 21 cm signal at cosmic dawn,''
Phys. Rev. D \textbf{99}, no.12, 123511 (2019)
%doi:10.1103/PhysRevD.99.123511
[arXiv:1809.01139 [astro-ph.CO]].
%22 citations counted in INSPIRE as of 12 Aug 2021



%\cite{Munoz:2018pzp}
\bibitem{Munoz:2018pzp}
J.~B.~Mu\~noz and A.~Loeb,
``A small amount of mini-charged dark matter could cool the baryons in the early Universe,''
Nature \textbf{557}, no.7707, 684 (2018)
%doi:10.1038/s41586-018-0151-x
[arXiv:1802.10094 [astro-ph.CO]].
%139 citations counted in INSPIRE as of 28 Apr 2020

%\cite{Fialkov:2018xre}
\bibitem{Fialkov:2018xre}
A.~Fialkov, R.~Barkana and A.~Cohen,
``Constraining Baryon--Dark Matter Scattering with the Cosmic Dawn 21-cm Signal,''
Phys. Rev. Lett. \textbf{121}, 011101 (2018)
%doi:10.1103/PhysRevLett.121.011101
[arXiv:1802.10577 [astro-ph.CO]].
%110 citations counted in INSPIRE as of 09 Aug 2021


%\cite{Berlin:2018sjs}
\bibitem{Berlin:2018sjs}
A.~Berlin, D.~Hooper, G.~Krnjaic and S.~D.~McDermott,
``Severely Constraining Dark Matter Interpretations of the 21-cm Anomaly,''
Phys. Rev. Lett. \textbf{121}, no.1, 011102 (2018)
%doi:10.1103/PhysRevLett.121.011102
[arXiv:1803.02804 [hep-ph]].
%139 citations counted in INSPIRE as of 27 May 2020

%\cite{Barkana:2018qrx}
\bibitem{Barkana:2018qrx}
R.~Barkana, N.~J.~Outmezguine, D.~Redigolo and T.~Volansky,
``Strong constraints on light dark matter interpretation of the EDGES signal,''
Phys. Rev. D \textbf{98}, no.10, 103005 (2018)
%doi:10.1103/PhysRevD.98.103005
[arXiv:1803.03091 [hep-ph]].
%127 citations counted in INSPIRE as of 27 May 2020

%\cite{Barkana:2018lgd}
\bibitem{Barkana:2018lgd}
R.~Barkana,
``Possible interaction between baryons and dark-matter particles revealed by the first stars,''
Nature \textbf{555}, no.7694, 71-74 (2018)
%doi:10.1038/nature25791
[arXiv:1803.06698 [astro-ph.CO]].
%212 citations counted in INSPIRE as of 29 Apr 2020

%\cite{Slatyer:2018aqg}
\bibitem{Slatyer:2018aqg}
T.~R.~Slatyer and C.~Wu,
``Early-Universe constraints on dark matter-baryon scattering and their implications for a global 21 cm signal,''
Phys. Rev. D \textbf{98}, no.2, 023013 (2018)
%doi:10.1103/PhysRevD.98.023013
[arXiv:1803.09734 [astro-ph.CO]].


%\cite{Jia:2018csj}
\bibitem{Jia:2018csj}
L.~B.~Jia,
``Dark photon portal dark matter with the 21-cm anomaly,''
Eur. Phys. J. C \textbf{79}, no.1, 80 (2019)
%doi:10.1140/epjc/s10052-019-6542-9
[arXiv:1804.07934 [hep-ph]].
%20 citations counted in INSPIRE as of 27 Oct 2021


%\cite{Houston:2018vrf}
\bibitem{Houston:2018vrf}
N.~Houston, C.~Li, T.~Li, Q.~Yang and X.~Zhang,
``Natural Explanation for 21 cm Absorption Signals via Axion-Induced Cooling,''
Phys. Rev. Lett. \textbf{121}, no.11, 111301 (2018)
%doi:10.1103/PhysRevLett.121.111301
[arXiv:1805.04426 [hep-ph]].
%27 citations counted in INSPIRE as of 27 Oct 2021

%\cite{Sikivie:2018tml}
\bibitem{Sikivie:2018tml}
P.~Sikivie,
%``Axion dark matter and the 21-cm signal,''
Phys. Dark Univ. \textbf{24}, 100289 (2019)
doi:10.1016/j.dark.2019.100289
[arXiv:1805.05577 [astro-ph.CO]].
%23 citations counted in INSPIRE as of 27 Oct 2021


%\cite{Kovetz:2018zan}
\bibitem{Kovetz:2018zan}
E.~D.~Kovetz, V.~Poulin, V.~Gluscevic, K.~K.~Boddy, R.~Barkana and M.~Kamionkowski,
``Tighter limits on dark matter explanations of the anomalous EDGES 21 cm signal,''
Phys. Rev. D \textbf{98}, no.10, 103529 (2018)
%doi:10.1103/PhysRevD.98.103529
[arXiv:1807.11482 [astro-ph.CO]].
%63 citations counted in INSPIRE as of 27 May 2020

%\cite{Boddy:2018wzy}
\bibitem{Boddy:2018wzy}
K.~K.~Boddy, V.~Gluscevic, V.~Poulin, E.~D.~Kovetz, M.~Kamionkowski and R.~Barkana,
``Critical assessment of CMB limits on dark matter-baryon scattering: New treatment of the relative bulk velocity,''
Phys. Rev. D \textbf{98}, no.12, 123506 (2018)
%doi:10.1103/PhysRevD.98.123506
[arXiv:1808.00001 [astro-ph.CO]].
%41 citations counted in INSPIRE as of 27 May 2020


%\cite{Creque-Sarbinowski:2019mcm}
\bibitem{Creque-Sarbinowski:2019mcm}
C.~Creque-Sarbinowski, L.~Ji, E.~D.~Kovetz and M.~Kamionkowski,
``Direct millicharged dark matter cannot explain the EDGES signal,''
Phys. Rev. D \textbf{100}, no.2, 023528 (2019)
%doi:10.1103/PhysRevD.100.023528
[arXiv:1903.09154 [astro-ph.CO]].
%8 citations counted in INSPIRE as of 07 May 2020


%\cite{Liu:2019knx}
\bibitem{Liu:2019knx}
H.~Liu, N.~J.~Outmezguine, D.~Redigolo and T.~Volansky,
``Reviving Millicharged Dark Matter for 21-cm Cosmology,''
Phys. Rev. D \textbf{100}, no.12, 123011 (2019)
%doi:10.1103/PhysRevD.100.123011
[arXiv:1908.06986 [hep-ph]].
%9 citations counted in INSPIRE as of 07 May 2020



%\cite{Li:2018kzs}
\bibitem{Li:2018kzs}
C.~Li and Y.~F.~Cai,
``Searching for the Dark Force with 21-cm Spectrum in Light of EDGES,''
Phys. Lett. B \textbf{788}, 70-75 (2019)
%doi:10.1016/j.physletb.2018.11.011
[arXiv:1804.04816 [astro-ph.CO]].
%21 citations counted in INSPIRE as of 12 Aug 2021


%\cite{Prinz:1998ua}
\bibitem{Prinz:1998ua}
A.~A.~Prinz, R.~Baggs, J.~Ballam, S.~Ecklund, C.~Fertig, J.~A.~Jaros, K.~Kase, A.~Kulikov, W.~G.~J.~Langeveld and R.~Leonard, \textit{et al.}
``Search for millicharged particles at SLAC,''
Phys. Rev. Lett. \textbf{81}, 1175-1178 (1998)
%doi:10.1103/PhysRevLett.81.1175
[arXiv:hep-ex/9804008 [hep-ex]].
%173 citations counted in INSPIRE as of 09 Aug 2021


%\cite{Dubovsky:2003yn}
\bibitem{Dubovsky:2003yn}
S.~Dubovsky, D.~Gorbunov and G.~Rubtsov,
``Narrowing the window for millicharged particles by CMB anisotropy,''
JETP Lett. \textbf{79}, 1-5 (2004)
%doi:10.1134/1.1675909
[arXiv:hep-ph/0311189 [hep-ph]].
%109 citations counted in INSPIRE as of 27 May 2020


%\cite{Dolgov:2013una}
\bibitem{Dolgov:2013una}
A.~Dolgov, S.~Dubovsky, G.~Rubtsov and I.~Tkachev,
``Constraints on millicharged particles from Planck data,''
Phys. Rev. D \textbf{88}, no.11, 117701 (2013)
%doi:10.1103/PhysRevD.88.117701
[arXiv:1310.2376 [hep-ph]].
%88 citations counted in INSPIRE as of 27 May 2020


%\cite{dePutter:2018xte}
\bibitem{dePutter:2018xte}
R.~de Putter, O.~Doré, J.~Gleyzes, D.~Green and J.~Meyers,
``Dark Matter Interactions, Helium, and the Cosmic Microwave Background,''
Phys. Rev. Lett. \textbf{122}, no.4, 041301 (2019)
%doi:10.1103/PhysRevLett.122.041301
[arXiv:1805.11616 [astro-ph.CO]].
%11 citations counted in INSPIRE as of 27 May 2020


%\cite{Chang:2018rso}
\bibitem{Chang:2018rso}
J.~H.~Chang, R.~Essig and S.~D.~McDermott,
``Supernova 1987A Constraints on Sub-GeV Dark Sectors, Millicharged Particles, the QCD Axion, and an Axion-like Particle,''
JHEP \textbf{09}, 051 (2018)
%doi:10.1007/JHEP09(2018)051
[arXiv:1803.00993 [hep-ph]].
%141 citations counted in INSPIRE as of 27 May 2020


%\cite{Boehm:2013jpa}
\bibitem{Boehm:2013jpa}
C.~Boehm, M.~J.~Dolan and C.~McCabe,
``A Lower Bound on the Mass of Cold Thermal Dark Matter from Planck,''
JCAP \textbf{08}, 041 (2013)
%doi:10.1088/1475-7516/2013/08/041
[arXiv:1303.6270 [hep-ph]].
%147 citations counted in INSPIRE as of 27 May 2020

%\cite{Depta:2019lbe}
\bibitem{Depta:2019lbe}
P.~F.~Depta, M.~Hufnagel, K.~Schmidt-Hoberg and S.~Wild,
``BBN constraints on the annihilation of MeV-scale dark matter,''
JCAP \textbf{04}, 029 (2019)
%doi:10.1088/1475-7516/2019/04/029
[arXiv:1901.06944 [hep-ph]].
%50 citations counted in INSPIRE as of 26 Oct 2021

%\cite{Kors:2005uz}
\bibitem{Kors:2005uz}
B.~Kors and P.~Nath,
``Aspects of the Stueckelberg extension,''
JHEP \textbf{07}, 069 (2005)
%doi:10.1088/1126-6708/2005/07/069
[arXiv:hep-ph/0503208 [hep-ph]].
%144 citations counted in INSPIRE as of 14 May 2020


%\cite{Feldman:2006ce}
\bibitem{Feldman:2006ce}
D.~Feldman, Z.~Liu and P.~Nath,
``Probing a very narrow Z-prime boson with CDF and D0 data,''
Phys. Rev. Lett. \textbf{97}, 021801 (2006)
%doi:10.1103/PhysRevLett.97.021801
[arXiv:hep-ph/0603039 [hep-ph]].
%94 citations counted in INSPIRE as of 14 May 2020


%\cite{Feldman:2006wb}
\bibitem{Feldman:2006wb}
D.~Feldman, Z.~Liu and P.~Nath,
``The Stueckelberg $Z$ Prime at the LHC: Discovery Potential, Signature Spaces and Model Discrimination,''
JHEP \textbf{11}, 007 (2006)
%doi:10.1088/1126-6708/2006/11/007
[arXiv:hep-ph/0606294 [hep-ph]].
%97 citations counted in INSPIRE as of 14 May 2020


%\cite{Feldman:2007wj}
\bibitem{Feldman:2007wj}
D.~Feldman, Z.~Liu and P.~Nath,
``The Stueckelberg Z-prime Extension with Kinetic Mixing and Milli-Charged Dark Matter From the Hidden Sector,''
Phys. Rev. D \textbf{75}, 115001 (2007)
%doi:10.1103/PhysRevD.75.115001
[arXiv:hep-ph/0702123 [hep-ph]].
%268 citations counted in INSPIRE as of 14 May 2020


%\cite{Feldman:2009wv}
\bibitem{Feldman:2009wv}
D.~Feldman, Z.~Liu, P.~Nath and B.~D.~Nelson,
``Explaining PAMELA and WMAP data through Coannihilations in Extended SUGRA with Collider Implications,''
Phys. Rev. D \textbf{80}, 075001 (2009)
%doi:10.1103/PhysRevD.80.075001
[arXiv:0907.5392 [hep-ph]].
%53 citations counted in INSPIRE as of 14 May 2020


%\cite{Du:2019mlc}
\bibitem{Du:2019mlc}
M.~Du, Z.~Liu and V.~Tran,
``Enhanced Long-Lived Dark Photon Signals at the LHC,''
JHEP \textbf{05}, 055 (2020)
%doi:10.1007/JHEP05(2020)055
[arXiv:1912.00422 [hep-ph]].
%4 citations counted in INSPIRE as of 24 Oct 2021


%\cite{Cline:2014dwa}
\bibitem{Cline:2014dwa}
J.~M.~Cline, G.~Dupuis, Z.~Liu and W.~Xue,
``The windows for kinetically mixed Z'-mediated dark matter and the galactic center gamma ray excess,''
JHEP \textbf{08}, 131 (2014)
%doi:10.1007/JHEP08(2014)131
[arXiv:1405.7691 [hep-ph]].
%114 citations counted in INSPIRE as of 24 Oct 2021


%\cite{Kolb:1990vq}
\bibitem{Kolb:1990vq}
E.~W.~Kolb and M.~S.~Turner,
``The Early Universe,''
Front. Phys. \textbf{69}, 1-547 (1990)
%1885 citations counted in INSPIRE as of 29 Jul 2020

%\cite{Ade:2015xua}
\bibitem{Ade:2015xua}
P.~Ade \textit{et al.} [Planck],
``Planck 2015 results. XIII. Cosmological parameters,''
Astron. Astrophys. \textbf{594}, A13 (2016)
%doi:10.1051/0004-6361/201525830
[arXiv:1502.01589 [astro-ph.CO]].
%8625 citations counted in INSPIRE as of 14 May 2020


%\cite{Munoz:2015bca}
\bibitem{Munoz:2015bca}
J.~B.~Mu\~noz, E.~D.~Kovetz and Y.~Ali-Ha\"imoud,
``Heating of Baryons due to Scattering with Dark Matter During the Dark Ages,''
Phys. Rev. D \textbf{92}, no.8, 083528 (2015)
%doi:10.1103/PhysRevD.92.083528
[arXiv:1509.00029 [astro-ph.CO]].
%69 citations counted in INSPIRE as of 28 Apr 2020


%\cite{Seager:1999km}
\bibitem{Seager:1999km}
S.~Seager, D.~D.~Sasselov and D.~Scott,
``How exactly did the universe become neutral?,''
Astrophys. J. Suppl. \textbf{128}, 407-430 (2000)
%doi:10.1086/313388
[arXiv:astro-ph/9912182 [astro-ph]].
%288 citations counted in INSPIRE as of 29 Apr 2020


%\cite{Peebles:1994xt}
\bibitem{Peebles:1994xt}
P.~J.~E.~Peebles,
``Principles of physical cosmology,''
%71 citations counted in INSPIRE as of 31 Jul 2020


%\cite{AliHaimoud:2010dx}
\bibitem{AliHaimoud:2010dx}
Y.~Ali-Haimoud and C.~M.~Hirata,
``HyRec: A fast and highly accurate primordial hydrogen and helium recombination code,''
Phys. Rev. D \textbf{83}, 043513 (2011)
%doi:10.1103/PhysRevD.83.043513
[arXiv:1011.3758 [astro-ph.CO]].
%118 citations counted in INSPIRE as of 29 Apr 2020





%\cite{McGaugh:2018dfk}
\bibitem{McGaugh:2018dfk}
S.~S.~McGaugh,
``Predictions for the Sky-Averaged Depth of the 21 cm Absorption Signal at High Redshift in Cosmologies with and without Nonbaryonic Cold Dark Matter,''
Phys. Rev. Lett. \textbf{121}, no.8, 081305 (2018)
%doi:10.1103/PhysRevLett.121.081305
[arXiv:1808.02532 [astro-ph.CO]].
%2 citations counted in INSPIRE as of 22 May 2020


%\cite{CMS:2012xi}
\bibitem{CMS:2012xi}
S.~Chatrchyan \textit{et al.} [CMS],
``Search for Fractionally Charged Particles in $pp$ Collisions at $\sqrt{s}=7$ TeV,''
Phys. Rev. D \textbf{87}, no.9, 092008 (2013)
%doi:10.1103/PhysRevD.87.092008
[arXiv:1210.2311 [hep-ex]].
%48 citations counted in INSPIRE as of 01 Jun 2020

%\cite{Magill:2018tbb}
\bibitem{Magill:2018tbb}
G.~Magill, R.~Plestid, M.~Pospelov and Y.~D.~Tsai,
``Millicharged particles in neutrino experiments,''
Phys. Rev. Lett. \textbf{122}, no.7, 071801 (2019)
%doi:10.1103/PhysRevLett.122.071801
[arXiv:1806.03310 [hep-ph]].
%38 citations counted in INSPIRE as of 01 Jun 2020


%\cite{ArgoNeuT:2019ckq}
\bibitem{ArgoNeuT:2019ckq}
R.~Acciarri \textit{et al.} [ArgoNeuT],
``Improved Limits on Millicharged Particles Using the ArgoNeuT Experiment at Fermilab,''
Phys. Rev. Lett. \textbf{124}, no.13, 131801 (2020)
%doi:10.1103/PhysRevLett.124.131801
[arXiv:1911.07996 [hep-ex]].
%27 citations counted in INSPIRE as of 29 Aug 2021

%\cite{Ball:2020dnx}
\bibitem{Ball:2020dnx}
A.~Ball, G.~Beauregard, J.~Brooke, C.~Campagnari, M.~Carrigan, M.~Citron, J.~De La Haye, A.~De Roeck, Y.~Elskens and R.~E.~Franco, \textit{et al.}
``Search for millicharged particles in proton-proton collisions at $\sqrt{s} = 13$ TeV,''
Phys. Rev. D \textbf{102}, no.3, 032002 (2020)
%doi:10.1103/PhysRevD.102.032002
[arXiv:2005.06518 [hep-ex]].
%16 citations counted in INSPIRE as of 29 Aug 2021


%\cite{Davidson:2000hf}
\bibitem{Davidson:2000hf}
S.~Davidson, S.~Hannestad and G.~Raffelt,
``Updated bounds on millicharged particles,''
JHEP \textbf{05}, 003 (2000)
%doi:10.1088/1126-6708/2000/05/003
[arXiv:hep-ph/0001179 [hep-ph]].
%337 citations counted in INSPIRE as of 01 Jun 2020


%\cite{Badertscher:2006fm}
\bibitem{Badertscher:2006fm}
A.~Badertscher, P.~Crivelli, W.~Fetscher, U.~Gendotti, S.~Gninenko, V.~Postoev, A.~Rubbia, V.~Samoylenko and D.~Sillou,
``An Improved Limit on Invisible Decays of Positronium,''
Phys. Rev. D \textbf{75}, 032004 (2007)
%doi:10.1103/PhysRevD.75.032004
[arXiv:hep-ex/0609059 [hep-ex]].
%124 citations counted in INSPIRE as of 01 Jun 2020


%\cite{Emken:2019tni}
\bibitem{Emken:2019tni}
T.~Emken, R.~Essig, C.~Kouvaris and M.~Sholapurkar,
``Direct Detection of Strongly Interacting Sub-GeV Dark Matter via Electron Recoils,''
JCAP \textbf{09}, 070 (2019)
%doi:10.1088/1475-7516/2019/09/070
[arXiv:1905.06348 [hep-ph]].
%24 citations counted in INSPIRE as of 02 Jun 2020

%\cite{Mahdawi:2018euy}
\bibitem{Mahdawi:2018euy}
M.~S.~Mahdawi and G.~R.~Farrar,
``Constraints on Dark Matter with a moderately large and velocity-dependent DM-nucleon cross-section,''
JCAP \textbf{10}, 007 (2018)
%doi:10.1088/1475-7516/2018/10/007
[arXiv:1804.03073 [hep-ph]].
%41 citations counted in INSPIRE as of 04 Sep 2021


%\cite{Fabbrichesi:2020wbt}
\bibitem{Fabbrichesi:2020wbt}
M.~Fabbrichesi, E.~Gabrielli and G.~Lanfranchi,
``The Dark Photon,''
[arXiv:2005.01515 [hep-ph]].
%1 citations counted in INSPIRE as of 26 May 2020


%\cite{Dodelson:2003ft}
\bibitem{Dodelson:2003ft}
S.~Dodelson,
``Modern Cosmology''. 
%544 citations counted in INSPIRE as of 31 Jul 2020

%\cite{McDermott:2010pa}
\bibitem{McDermott:2010pa}
S.~D.~McDermott, H.~Yu and K.~M.~Zurek,
``Turning off the Lights: How Dark is Dark Matter?,''
Phys. Rev. D \textbf{83}, 063509 (2011)
%doi:10.1103/PhysRevD.83.063509
[arXiv:1011.2907 [hep-ph]].
%149 citations counted in INSPIRE as of 28 Apr 2020




% Please avoid comments such as "For a review'', "For some examples",
% "and references therein" or move them in the text. In general,
% please leave only references in the bibliography and move all
% accessory text in footnotes.

% Also, please have only one work for each \bibitem.





\end{thebibliography}
\end{document}